# EXTENDING THE APPLICATION OF DYNAMIC BAYESIAN NETWORKS IN CALCULATING MARKET RISK: STANDARD AND STRESSED EXPECTED SHORTFALL

Eden Gross, School of Statistics and Actuarial Science, University of the Witwatersrand, Johannesburg, South Africa. eden.gross@wits.ac.za ORCiD: 0000-0002-7647-8890

Ryan Kruger, Department of Finance and Tax, University of Cape Town, Cape Town, South Africa. ryan.kruger@uct.ac.za

Francois Toerien, Department of Finance and Tax, University of Cape Town, Cape Town, South Africa. francois.toerien@uct.ac.za ORCiD: 0000-00002-6051-1094

## ABSTRACT

In the last five years, expected shortfall (ES) and stressed ES (SES) have become key required regulatory measures of market risk in the banking sector, especially following events such as the global financial crisis. Thus, finding ways to optimize their estimation is of great importance. We extend the application of dynamic Bayesian networks (DBNs) to the estimation of 10-day 97.5% ES and stressed ES, building on prior work applying DBNs to value at risk. Using the S&P 500 index as a proxy for the equities trading desk of a US bank, we compare the performance of three DBN structure-learning algorithms with several traditional market risk models, using either the normal or the skewed Student's t return distributions. Backtesting shows that all models fail to produce statistically accurate ES and SES forecasts at the 2.5% level, reflecting the difficulty of modeling extreme tail behavior. For ES, the EGARCH(1,1) model (normal) produces the most accurate forecasts, while, for SES, the GARCH(1,1) model (normal) performs best. All distribution-dependent models deteriorate substantially when using the skewed Student's t distribution. The DBNs perform comparably to the historical simulation model, but their contribution to tail prediction is limited by the small weight assigned to their one-day-ahead forecasts within the return distribution. Future research should examine weighting schemes that enhance the influence of forward-looking DBN forecasts on tail risk estimation.

**Keywords:**

Market risk forecasting; Basel Accords; Risk management; Equities

**JEL codes:**

G17, G21, G32, C51, C53

**Statements and Declarations:**

*Competing Interests and Funding*
The authors declare that they are not aware of any competing interests that exist with regards to this research. This research did not receive any specific funding from any funding agency in the public, commercial, or not-for-profit sectors.

*Declaration of the use of generative artificial intelligence in scientific writing*
During the preparation of this work, the author(s) used Copilot to improve the flow and grammar of the manuscript. After using this tool/service, the author(s) reviewed and edited the content as needed and take(s) full responsibility for the content of the published article.

# 1.	Introduction

Before the 2008 global financial crisis, banks primarily relied on value at risk (VaR) as the standard measure for quantifying market risk – estimating the expected loss of portfolio value over a given time horizon at a specified confidence level due to changes in market prices (Alexander, 2008). While VaR provided a simple and widely-adopted framework, the 2008 global financial crisis revealed its limitations, particularly its inability to capture extreme tail risks and losses beyond the chosen confidence threshold. In response, regulators, guided by the Basel Committee on Banking Supervision's (BCBS's) Basel III and, later, Basel IV, introduced expected shortfall (ES) as a complementary and, in many cases, replacement measure. ES provides a more comprehensive view of market risk by accounting for the average tail loss beyond the VaR figure, thereby offering a more complete assessment of a bank's exposure during periods of severe market stress. In addition to the BCBS's move from VaR and ES, it also introduced stressed ES (SES) as an augmenting market risk metric. This measure is calculated similarly to ES, except that it uses a stressed period as its calibration period, where this stressed period must include the 2008 global financial crisis when determined (Basel Committee on Banking Supervision, 2023). The BCBS allows the use of internal models to calculate these forecasts (see Basel Committee on Banking Supervision, 1996), but requires both ES and SES to be calculated directly as 10-day 97.5% ES and 10-day 97.5% SES forecasts.

The increased computing power and prevalence of machine learning algorithms have resulted in the increased feasibility of applying such algorithms to existing issues, including in the area of market risk management. Bayesian networks (BNs) are one such application. These networks can be used to learn the causal relationships between various variables that influence each other, resulting in a probability density function (PDF) of the target variable. When applied to the management of market risk of an equities trading desk at a bank, as is the focus of this study, BNs can be used to learn the causal relationships between economic and financial variables and a proxy for the United States (US) bank's equities trading desk's returns, where we use the Standard and Poor's (S&P) 500 index as that proxy. These relationships can then be used to produce forward-looking return predictions and, in turn, more accurate tail risk metric forecasts, such as VaR, ES, and their stressed counterparts.

The use of BNs to calculate market risk was proposed by early studies, such as those of Shenoy and Shenoy (2000) and Demirer, Mau, and Shenoy, (2006). More recently, Apps (2020) developed a simplified BN methodology to predict the directional move of a portfolio's returns and its impacts on VaR, which was, subsequently, expanded upon to calculate 10-day 99% VaR and SVaR figures by Gross, Kruger, and Toerien (2025). Our study expands the work presented by Gross, et al., (2025) by introducing BNs to the calculation of 10-day 97.5% ES and SES forecasts.

The use and incorporation of BNs in producing market risk forecasts may overcome some of the challenges experienced in quantifying and managing market risk. First, due to the causal relationships learned within the network and their frequent updating, the BN may result in market risk metric forecasts that are more sensitive to changes in market volatility, thereby overcoming the slow adaptation displayed by currently used models such as the historical simulation model (Pérignon & Smith, 2010). Second, this frequent updating using modelled causal relationships may lead to market risk metric forecasts that are more in line with market movements, thereby reducing the underestimation of risk when using 10-day forecasts (Gross, et al., 2025). Third, the use of BNs will bring financial risk management techniques to the modern age, by initiating an exploration of forward-looking forecasting techniques beyond the tools available in the toolbox of the traditional risk manager at a bank.

We follow the methodological foundation detailed by Gross, et al., (2025), by using dynamic BNs (DBNs) to forecast these market risk metrics, and simply refer to these as BNs in the remainder of this paper. DBNs, as opposed to static BNs, account for the interactions of variables both between periods and within a period (Friedman, Murphy, & Russell, 2013; Dagum, Galper, & Horvitz, 1992), thereby offering structure learning that fits the nature of financial time series data better. Further, we use the stressed period determination methodology presented by Gross, et al., (2025), whereby the stressed period is determined using a rolling period methodology as the amalgamation of the most severe return days of the S&P 500 index over the study period, where such days are not necessarily consecutive days. We believe that this methodology offers the most conservative approach to stressed market risk management, especially in the highly-regulated banking industry.

This study contributes to the literature in several ways. First, for 10-day 99% VaR and SVaR, we provide a comprehensive assessment and evaluation of traditional market risk models, focusing here on forecasting ES and SES. This, too, is done over an extended out-of-sample period (1991-2020). This is incredibly important considering that the BCBS requires banks to use ES and SES since the introduction of the Basel III accord, and since the literature covering either ES or SES is very scarce relative to that of the well-established research on VaR. Second, we do this comparison using the normal and skewed Student's t distributions, adding robustness to the analyses provided in this study. While the normal distribution is generally used when producing VaR forecasts and, therefore, is used when producing ES forecasts, we also test the skewed Student's t distribution, given that its skewness and heavier tails may capture equity returns more accurately than the normal distribution. While Gross, et al., (2025) find that the skewed Student's t distribution does not produce more accurate VaR, its application to the tail beyond VaR, as is the case for ES, may help in definitively determining the suitability of this distribution to the tail as a whole. Third, we extend the BN-related analyses and conclusions presented by Gross, et al., (2025) to the rest of the tail, i.e., to ES and SES, and provide conclusions that are relevant to academics and practitioners alike surrounding the application of BNs to their complete arsenal of tail-based market risk metrics.

The remainder of this paper is structured as follows. Section 2 provides a literature review and the technical context underpinning this paper. Section 3 then discusses the data and methodology, while Section 4 discusses and analyses our results. Section 5 concludes this paper.

## 2.    Literature Review and Technical Context

ES is often regarded as a substitute or a complement risk management measure to the VaR measure, since VaR does not quantify the risk in the tail of the profit and loss distribution. Indeed, this is one of the primary reasons for the BCBS's shift from VaR to ES (Basel Committee on Banking Supervision, 2019a). Moreover, ES is a coherent risk measure, and some authors regard it as a better risk metric relative to the non-coherent VaR (see, for example, Acerbi and Tasche, 2002).

ES, also known as conditional VaR or expected tail loss, was introduced by Artzner, Delbaen, Eber, and Heath (1997) as the expected loss of a portfolio given the exceedance of the VaR forecast (Yamai & Yoshiba, 2005). Hence, by definition, ES is a measure explaining the tail risk beyond the VaR forecast, and requires the calculation of VaR first.

Let $X$ be the random variable representing the profit and loss account of the equities trading desk of a bank. We define the $\alpha$ quantile $h$-day profit and loss value at time $t$ to be $x_{ht,\alpha}$, such that

$$\Pr[X_{ht} < x_{ht,\alpha}] < \alpha \qquad (1)$$

VaR at time $t$ is then calculated as follows.

$$VaR_{ht,\alpha} = -x_{ht,\alpha} \times P_t \qquad (2)$$

where $P_t$ represents the total portfolio value at time $t$ (Alexander, 2008).

The definition of ES as a coherent risk measure, adapted from Acerbi and Tasche (2002), is as follows: Consider the same profit and loss random variable $X$, as defined for VaR in the explanation preceding Equation (1), again with a probability level of $100\alpha\%$, specified over the time horizon $t$. The expected shortfall corresponding to $100\alpha\%$ is then equal to

$$ES^{(\alpha)}(X) = -\frac{1}{\alpha}\Big[E\Big[X\mathbf{1}_{\{X \le x^{(\alpha)}\}}\Big] - x^{(\alpha)}\big(F\big(x^{(\alpha)}\big) - \alpha\big)\Big] \qquad (3)$$

where $\mathbf{1}_{\{X \le x^{(\alpha)}\}}$ is an indicator variable equal to 1 if $X \le x^{(\alpha)}$ or 0 if $X > x^{(\alpha)}$; and $x^{(\alpha)} = \sup\{x \mid \Pr[X \le x] \le \alpha\}$.

We can also define ES explicitly in terms of VaR, as it is the expected loss given the exceedance of the VaR forecast. Hence, given a profit and loss distribution $X$ and a probability level $100\alpha\%$, ES can be defined as the average loss exceeding the $100\alpha\%$ $h$-day VaR forecast, mathematically defined as follows.

$$ES_{ht,\alpha} = E\big[X \big| X \ge VaR_{ht,\alpha}\big] = ES^{(\alpha)} \qquad (4)$$

We consider several widely-used models to forecast ES and SES and refer to these as traditional models. These traditional models include the historical simulation, the delta-normal, the autoregressive conditional heteroscedasticity(1) (ARCH(1)), the generalized ARCH(1,1) (GARCH(1,1)) model, and the exponential GARCH(1,1) (EGARCH(1,1)) model, along with the RiskMetrics model.

The ARCH(1) model is defined as follows.

$$\sigma_{t+1}^2 = \omega + \alpha\varepsilon_t^2, \qquad \omega > 0, \qquad \alpha \ge 0, \qquad \varepsilon_t = \sigma_t z_t \qquad (5)$$

where $\sigma_t$ is the return volatility; the $z_t$ is a series of independently and identically distributed standard normal random variables (Bollerslev, 2007). The GARCH(1,1) model similarly makes use of squared residuals, however, the contribution of squared residuals at time $\tau$ diminishes asymptotically to zero as time $\tau$ moves further away from the current time $t$, i.e., as the contribution moves further into the past (Engle, 2001). The mode is specified as follows.

$$\sigma_{t+1}^2 = \omega + \alpha\sigma_t^2\varepsilon_t^2 + \beta\sigma_t^2, \qquad \alpha, \beta, \omega > 0, \qquad \alpha + \beta < 1 \tag{6}$$

where the process's long-term average is $\sqrt{\omega/[1 - (\alpha + \beta)]}$; and the $\varepsilon_t$ is the residual term of the time series. The EGARCH(1,1) model additionally accounts for leverage effects, which are not captured by the previous models, where:

$$\ln(\sigma_{t+1}^2) = \omega + \beta\ln(\sigma_t^2) + \alpha\left(\frac{|\varepsilon_t|}{\sigma_t} - E\left[\frac{|\varepsilon_t|}{\sigma_t}\right]\right) + \gamma \cdot \frac{\varepsilon_t}{\sigma_t} \tag{7}$$

Finally, RiskMetrics is an exponentially-weighted moving average model for volatility specified as follows.

$$\sigma_{t+1}^2 = \lambda\sigma_t^2 + (1 - \lambda)r_t^2 \tag{8}$$

where $\lambda = 0.94$ is the smoothing parameter; and $r_t$ is the previous day's return.

We calibrated the ARCH, GARCH, EGARCH, and RiskMetrics models using either the normal or skewed Student's t distribution as the underlying return distribution. The historical simulation and delta-normal models are distribution agnostic.

While there is substantial research on the performances of these traditional models in estimating VaR, the literature for ES is less prevalent, with much of it focusing on more complex theoretical models incorporating extreme value theory, filtering and quantile-regression (see, for example, Novales and Garcia-Jorcano, 2019 and Storti and Wang, 2022).

Žiković and Filer (2013) compare and rank the performances of various models when calculating VaR and ES for eight developed and eight emerging market equity indices over a ten-year period. While they tested a broader range of models than those examined here, they found that the worst performing models for both VaR and ES were the historical simulation and RiskMetrics models.

García-Risueño (2025) further highlight the shortcomings of the historical simulation model in estimating ES. The author fitted returns to heavy-tailed distributions (including the Student's t distribution, although not its skewed variant) and showed that they capture tail risk better than the normal distribution. Using these fitted distributions to generate synthetic data, García-Risueño (2025) reveals how the historical simulation model systematically underestimates ES compared to other parametric approaches, and strongly encourages regulators to limit the model's use.

A key process in determining the performance of a market risk forecasting model in the BCBS framework is backtesting, whereby we evaluate a model's performance by comparing its daily forecasts to the actual historical returns (Basel Committee on Banking Supervision, 2019b). We record a breach (or exceedance, exception) if the daily loss exceeds the model's forecast, and evaluate the statistical accuracy of the model based on the actual number of breaches relative to the expected number of breaches. A drawback of ES as a risk measure is that it is not elicitable (Gneiting, 2011) as VaR is, meaning that it is not easily ranked in performance across different models. In fact, Diebold, Gunther, and Tay (1998) conclude that ES cannot be backtested at all. However, some authors, such as Acerbi and Székely (2014), provide backtests which do not rely on the elicitability of a risk measure, or backtests that rely on the joint elicitability of VaR and ES.

One of the most commonly used approaches to backtest ES is the traffic light test of the BCBS (Chen, 2018). The traffic light test uses a cumulative probability approach to classify models into one of three 'zones', namely a 'Green zone', a 'Yellow zone', and a 'Red zone'. In addition to employing the traffic light test, we also employ more theoretically rigorous backtests, namely the conditional backtest (Acerbi & Székely, 2014), the minimally biased backtest (Acerbi & Székely, 2017), and the Du-Escanciano independence test (Du & Escanciano, 2017).

The backtests introduced by Acerbi and Székely (2014) assume that the independence of any ES breach has been tested independently of the statistical backtests proposed (Acerbi & Székely, 2014). He, Kou, and Peng (2022) highlight that the three tests proposed by Acerbi and Székely (2014) are indirect backtests[1] of ES, meaning that the accuracy of the backtests deteriorates for larger banks, implying the larger bank's increased likelihood to under-report ES relative to smaller banks (He, et al., 2022). This may have regulatory consequences in real-world applications, although only in economies that require banks of all sizes to comply with the Basel Accords. This is not the case in the US, where only first-tier banks are required

---

[1] A backtest is said to be indirect if it either: (a) Examines whether the distribution (in full or in part, e.g., the tail) or its properties correspond to the quantities of the true, yet unknown, distribution; or (b) Examines the backtestability of a collection of risk measures, which are elicitable collectively (He, et al., 2022).

to comply with the Basel Accords, while compliance for smaller banks is optional (Herring, 2007).

The application of BNs in market risk management is very limited. Early studies by Shenoy and Shenoy (2000) and Demirer, et al., (2006) have hypothesized the utility that can be derived from the network belief and causal updating capabilities of BNs in financial risk management, but few authors have applied BNs in this context. Apps (2020) used a BN and the assumption of multivariate normality to produce VaR forecasts based on directional changes in portfolio return (i.e., based on profit or loss). This was then expanded upon by Gross, et al., (2025) by presenting a complete top-down approach, as outlined by Demirer, et al., (2006), to forecast the returns of the S&P 500 index and to use these forecasts to calculate 10-day 99% VaR and SVaR forecasts. We now expand this study to include 10-day 97.5% ES and SES forecasts, thereby completing the BN risk management toolbox of academics and practitioners alike.

In doing so, we contribute to the literature in several areas. First, we provide a comprehensive comparison of the performances of traditional models when producing 10-day 97.5% ES and SES forecasts, noting the scarce literature surrounding these, especially in the context of banking regulation. Second, we do this over an extended out-of-sample period using either the normal or skewed Student's t distribution, where appropriate, to increase to robustness of our analyses and model evaluations. Third, we introduce the amalgamated stressed period construction approach of Gross, et al., (2025) to the context of SES. Last, we extend the ideas developed by Shenoy and Shenoy (2000) and Demirer, et al., (2006), as developed further and applied by Gross, et al., (2025) to the rest of the tail, i.e., S/ES, thereby extending the BN toolbox of academics and practitioners alike.

## 3.    Data and Methodology

We use daily closing values of the S&P 500 index from 15 March 1991 to 14 February 2020, obtained from the Bloomberg database, to forecast 7,286 out-of-sample 10-day 97.5% ES and SES forecasts using ten traditional models and three BNs. The S&P 500 index is used as a proxy for the equities trading desk of a US bank, following Gross, et al., (2025) and other studies (see, for example, Grinold, 1992), as the index tracks liquid and large market capitalization stocks in the US, believed to proxy the investments of the trading desk. The long out-of-sample period allows us to employ rolling period methodologies for both traditional and BN models, and covers approximately three business cycles in the US (National Bureau of Economic Research, n.d.). We collected the index's closing values for the entire out-of-sample period as well as for two preceding and sequential periods, each 1,264 days in length, as is usual in the literature and used by Gross, et al., (2025). The first of such periods is the initial training period for the BNs, discussed further later in this section, while the latter of such periods is the calibration period for the market risk models to facilitate market risk metric forecasting.

We calculated the S&P 500 index's daily returns, i.e., the bank's equities trading desk's profit and loss account, using the natural logarithm of the ratio of index prices (or levels) on consecutive days, i.e., the ratio of $P_t$ to $P_{t-1}$, as follows.

$$r_t = \ln\left(\frac{P_t}{P_{t-1}}\right) \tag{9}$$

The returns' descriptive statistics include an average return of 0.03%, a standard deviation of 1.10%, and a minimum and maximum of -9.47% and 10.96%, respectively. The return skewness coefficient was -0.28, while the return distribution had a kurtosis equal to 12.10.

When calculating the 10-day 97.5% S/ES forecasts, this study also follows the Basel Accords' guidance by calculating the 10-day 97.5% forecasts directly (i.e., without scaling) using a series of overlapping 10-day returns. As for SES, the methodology applied is the same as per the non-stressed version, except that a stressed period is used to forecast the stressed metric. The stressed period is the most severe period preceding the return's date, corresponding in length to the non-stressed calibration period, i.e., 1,264 days, for consistency.

The traditional models (historical simulation, delta-normal, ARCH, GARCH, EGARCH, and RiskMetrics) used the daily returns and the 1,264-day rolling calibration period to calculate the 7,286 out-of-sample 10-day 97.5% S/ES forecasts, using either the normal or skewed Student's t distribution, where appropriate. The autoregressive and RiskMetrics models maximized the log-likelihood using either distribution.

Following Gross, et al., (2025), we used the Peter and Clark (Stable) (PC (Stable)) algorithm, (Spirtes & Glymour, 1991), the max-min hill climbing (MMHC) algorithm, (Tsamardinos, Brown, & Aliferis, 2006), and the semi-interleaved HITON parents-children (SI-HITON-PC) algorithm (Aliferis, Tsamardinos, & Statnikov, 2003) as our BNs to produce the closing values of the S&P 500 index which, in turn, were used to produce the S/ES forecasts. These three algorithms learned the network structure (i.e., the causal relationships between and within variables over time) using a 1,264-day rolling training period, where the optimal structure for each training iteration was the one that minimized the Akaike information criterion (AIC). To facilitate the training of the networks, the forecasting of one-day-ahead closing values of the index, and to calculate both market risk metrics, Bloomberg data from 18 March 1981 were used, which affected the availability of a handful of variables for calibration (see below).

The BNs were trained on financial time series data of 41 macroeconomic and financial variables or proxies (detailed in Appendix A), identified through literature or expert judgment as potentially causally related to the S&P 500 index. Once again, our approach is the top-down framework outlined by Demirer, et al., (2006). We used the dbnR package, among other resources, to construct and learn the network structures for the DBN algorithms. Similarly to Gross, et al., (2025), and due to the algorithms' requirement for complete datasets, non-daily data were converted to daily by carrying forward the last known value. Echoing Gross, et al., (2025), this is a robust choice reflecting practical data usage, where only the latest value of a variable is available to the risk manager, forcing the risk manager to implicitly 'carry' this value until a new value is available or released. Further, the discussion of Gross, et al., (2025), surrounding data availability during the initial training period (up to a maximum of 10% of the data) applies to this study, too, and readers are encouraged to refer to Gross, et al., (2025) for further detail surrounding this and the advantages of the evaluation at the trading desk level.

We employ several backtesting models to assess the statistical accuracy of both the traditional and BN S/ES models. These backtests include the BCBS's regulatory traffic light test, which Gross, et al., (2025) find to be less of a backtest and more of a check to assess the excessiveness of model breaches, as well as the conditional backtest of Acerbi and Székely (2014), the minimally biased backtest of Acerbi and Székely (2017), and the Du-Escanciano backtest (Du & Escanciano, 2017). The conditional and minimally biased backtests' null hypotheses state that the ES forecasts are equal to the true ES values, where, for the conditional backtest, these are dependent on the existence of at least one VaR breach (Acerbi & Székely, 2014; 2017). The alternative hypotheses, on the other hand, state that the forecasts produced by the models underestimate the true ES values, where, for the minimally biased backtests, the reference is to at least some forecasts, rather than all, as is the case for the conditional backtest (Acerbi & Székely, 2014; 2017).

The conditional backtest (Acerbi & Székely, 2014) rests on the assumption that ES forecasts are produced and tested after their corresponding VaR forecasts have been produced and tested.

Using the definition of ES as a starting point, as stated in Equation (4), the following expectation can be derived.

$$E\left[\frac{X}{ES_{ht,\alpha}} + 1 \,\middle|\, X + VaR_{ht,\alpha} < 0\right] = 0 \qquad (10)$$

Since the null hypothesis states that $ES_{ht,\alpha} = \widehat{ES}_{ht,\alpha}$ (Acerbi & Székely, 2014), the test statistic, is as follows.

$$\bar{Z}_{CB}(X) = \frac{1}{N_{Breaches}} \sum_{t=1}^{N} \frac{X_t \mathbf{1}_{\{X \le x^{(\alpha)}\}}}{\widehat{ES}_{ht,\alpha}} + 1 \qquad (11)$$

where $X$ is a vector of profit and loss amounts; $\widehat{ES}_{ht,\alpha}$ is a vector of expected shortfall forecasts; $\mathbf{1}_{\{X \le x^{(\alpha)}\}}$ is as defined for Equation (3); and $N_{Breaches} = \sum_{t=1}^{N} \mathbf{1}_{\{X \le x^{(\alpha)}\}}$ (Acerbi & Székely, 2014), i.e., $N_{Breaches}$ is the number of VaR breaches observed.

Acerbi and Székely (2014) also propose an unconditional backtesting technique, but find subsequently that it displays a significant linear relationship to VaR forecasts (Acerbi & Székely, 2017), meaning that the incorrect calibration of the model to produce VaR forecasts may lead to an increased probability of producing either a Type I error or a Type II error. Instead, they propose the minimally biased backtest. The test statistic, $\bar{Z}_{MB}$, is as follows.

$$\bar{Z}_{MB}(X) = \frac{1}{N} \sum_{t=1}^{N} \widehat{ES}_{ht,\alpha} - \widehat{VaR}_{ht,\alpha} + \frac{(\widehat{VaR}_{ht,\alpha} + \widehat{ES}_{ht,\alpha})\mathbf{1}_{\{X \le x^{(\alpha)}\}}}{\alpha} \qquad (12)$$

where $X$ and $\widehat{ES}_{ht,\alpha}$ are, again, as defined for Equation (10); $\mathbf{1}_{\{X \le x^{(\alpha)}\}}$ is, again, as defined for Equation (3), and $\widehat{VaR}_{ht,\alpha}$ is as defined for Equation (2) (Acerbi & Székely, 2017). Note that this backtest is not conditioned on the existence of at least one VaR breach, unlike the conditional backtest.

For both backtests, under the null hypothesis, the expected values of the test statistics are zero (where, for the conditional backtest, this expectation is conditional on the corresponding number of VaR breaches being positive, i.e., $E[\bar{Z}_{CB}|N_{Breaches} > 0] = 0$), while under the alternative hypothesis, this expectation is negative (Acerbi & Székely, 2014; 2017). These expected values can be examined to evaluate the performance of the ES forecasting model as (any) negative values would indicate that the null hypothesis should be rejected for the model at hand, at the chosen confidence level.

We also employ the Du-Escanciano independence backtest for ES to complement any visual inspections of clustered ES breaches, as it is the ES-equivalent to Christoffersen's independence test for VaR forecasts. Du and Escanciano (2017) develop a Portmanteau Box-Pierce conditional backtest for the independence of ES breaches, i.e., a statistical test whose primary objective is to determine whether ES breaches cluster together.

Consider the conditional parametric distribution of the profit and loss account, conditional on some parameter $\Theta \in \mathbb{R}^p$, that unknown parameter being $\theta_0$. Du and Escanciano (2017) then show that the function of associated cumulative breaches, $H_t(\alpha, \theta_0)$, is as follows.

$$H_t(\alpha, \theta_0) = \frac{1}{\alpha}(\alpha - u_t(\theta_0))\mathbf{1}_{\{u_t(\theta_0) \leq \alpha\}} \tag{13}$$

where $u_t(\theta_0)$ is the conditional cumulative distribution function of the profit and loss account of a bank at time $t$, evaluated at the level $u$, conditional on the parameter $\theta_0$ and the filtration system $\mathcal{F}_t$, the set of which is termed by Du and Escanciano as generalised errors (Du & Escanciano, 2017).

The conditional test presented in Du and Escanciano (2017) is based on such cumulative breaches, as captured in Equation (13). This conditional test essentially tests whether the terms of the series $\{H_t(\alpha) - \alpha/2\}_{i=1}^{\infty}$ are uncorrelated. The null hypothesis states that, given the filtration system $\mathcal{F}_t$, the expected value of each term of this series is zero, i.e., $E[H_t(\alpha) - \alpha/2|\mathcal{F}_t] = 0$.

Consider the autocovariance function and autocorrelation function of $H_t(\alpha)$ for lag $j$ as $\gamma_{N,j}$ and $\rho_{N,j}$, where $\rho_j = \gamma_{N,j}/\gamma_{N,0}$ as usual. The Du-Escanciano conditional backtest statistic is, then, as follows.

$$\bar{Z}_{DE}(X, n) = N \sum_{j=1}^{n} \hat{\rho}_{T,j}^2 \tag{14}$$

where $N$ is the number of out-of-sample trading days; $n$ is the highest lag of autocorrelations, or 1 as default; and $\hat{\rho}_{N,j}$ is the estimate of $\rho_{N,j}$ as defined above (Du & Escanciano, 2017). Du and Escanciano (2017) show that this test statistic is asymptotically distributed as a chi-squared random variable with $n$ degree of freedom.

The backtesting techniques we employ can help us to determine whether the models used produce accurate S/ES forecasts. However, a crucial part in a bank's cost-benefit analysis of the choice of internal market risk model is the balance between breaches and excessive capital reserves, as the S/ES forecasts directly feed into the level of regulatory reserves the bank must hold. To assess the excess in capital held by the bank, as well as the proximity of the models' forecasts to the actual loss incurred by the equities trading desk of the generic US bank considered in this study, i.e., the efficiency of the models, we employ three forecasting error measures. These are the mean absolute error (MAE), root mean square error (RMSE), and mean absolute percentage error (MAPE). We use the symmetric MAPE (SMAPE) where the returns on the S&P 500 index are zero, as these invalidate the use of the MAPE. More efficient models will result in lower (cumulative) differences between the S/ES forecasts and actual index returns. Hence, lower values indicate more accurate models. These measures are crucial when few breaches are experienced or when models fail any of the backtests employed, as either would indicate a statistically inaccurate or conservative model, which would be the result of excessive capital held based on the conservative model forecasts.

## 4. Results

### 4.1. Expected Shortfall

The numbers of breaches observed for each of the 10-day 97.5% ES models used over the 7,286-trading-day out-of-sample period are summarized in Table 1, below. The historical simulation model, the primary model used by banks in the US (Pérignon & Smith, 2010), together with the three BNs, achieved three breaches, placing them as the models to yield the second fewest breaches, trailing only the EGARCH model using the skewed Student's t distribution as the underlying return distribution. The performances of the various models showed mixed results when replacing the normal distribution for the skewed Student's t distribution as the underlying statistical distribution. For example, the number of breaches observed for the GARCH and the RiskMetrics models increased (i.e., the models' performances deteriorated) when using the skewed Student's t distribution instead of the normal distribution, while the number of breaches observed for the EGARCH model decreased (i.e., the performance improved). The ARCH model achieved 19 breaches regardless of which underlying return distribution was used.

Table 1: Number of 10-day 97.5% Expected Shortfall Breaches

| Traditional Models | Normal Distribution | Skewed Student's t Distribution |
|---|---|---|
| *ARCH(1)* | 19 | 19 |
| *GARCH(1,1)* | 10 | 21 |
| *EGARCH(1,1)* | 8 | 2 |
| *RiskMetrics* | 11 | 18 |
| | | |
| *Historical Simulation* | 3 | |

| *Delta-Normal* | 7 |
| --- | --- |

**BN Models**

| *MMHC* | 3 |
| --- | --- |
| *PC (Stable)* | 3 |
| *SI-HITON-PC* | 3 |

The three BNs experienced three breaches each on the same dates as the historical simulation model (27 October 1997, 31 August 1998, and 29 September 2008). As was concluded for 10-day 99% VaR forecasting by Gross, et al., (2025), this observation means that banks can expect similar performance from any of the BN algorithms we employed in this study relative to the historical simulation model, where this performance would be in line with banks' preferences for low breaches that do not raise alarm with the regulator (McAleer & da Veiga, 2008).

All models obtain a 'Green zone' outcome using the BCBS's traffic light test, indicating that the forecasts produced are acceptable under this regulatory backtest, satisfying this regulatory requirement. This result is unsurprising given the low number of breaches relative to the total trading days in the out-of-sample period (even for the worst performing models).

The first ES-specific backtest we employed is the conditional backtest, whereby the 10-day 97.5% ES forecasts were evaluated together with their respective 10-day VaR forecasts. The backtest's results are shown in Table 2, below. Note that the conditional backtest could only be performed where at least one VaR breach (corresponding to the ES forecasts) was observed, rendering the backtest impossible to employ in the cases where there were no VaR breaches. In such a case, the backtest's result was captured as 'Cannot perform backtest' in Table 2. For the remaining models that can be tested (i.e., those with at least one corresponding VaR breach), this backtest rejects the null hypothesis at the 97.5% confidence level, meaning that the ES forecasts produced by the various models are statistically inaccurate.

Table 2: Conditional Backtest Results for the 10-day 97.5% Expected Shortfall Forecasts

| **Traditional Models** | **Normal Distribution** | **Skewed Student's t Distribution** |
| --- | --- | --- |
| *ARCH(1)* | Reject | Cannot perform backtest |
| *GARCH(1,1)* | Cannot perform backtest | Reject |
| *EGARCH(1,1)* | Cannot perform backtest | Reject |
| *RiskMetrics* | Reject | Reject |

| | |
|---|---|
| *Historical Simulation* | Reject |
| *Delta-Normal* | Reject |

**BN Models**

| | |
|---|---|
| *MMHC* | Cannot perform backtest |
| *PC (Stable)* | Cannot perform backtest |
| *SI-HITON-PC* | Cannot perform backtest |

Note: This table reports the results of the conditional backtest for banks' internal models based on the number of breaches of various 10-day expected shortfall (ES) models at the 97.5% confidence level using either the normal distribution or the skewed Student's t distribution as the underlying return distribution (where applicable) and the daily logged returns of the Standard & Poor's (S&P) 500 index from 15 March 1991 to 14 February 2020. Note that the historical simulation and the delta-normal models are distribution agnostic, meaning that the number of breaches experienced does not vary with the use of either distribution as the underlying return distribution. The test's null hypothesis states that the ES forecasts observed are the true ES figures. The test can only be carried out if there is at least one value at risk breach. The total number of ES forecasts for the study period was 7,286 per model. The Bayesian network (BN) learning algorithms considered were the max-min hill-climbing (MMHC) algorithm, the Peter and Clark Stable (PC (Stable)) algorithm, and the semi-interleaved HITON parents and children (SI-HITON-PC) algorithm.

Next, we employed the minimally biased backtest to the 10-day 97.5% ES forecasts. As for the conditional backtest, the null hypothesis states that the observed 10-day 97.5% ES forecasts are the true 10-day 97.5% ES forecasts. In contrast to the preceding backtest, however, this backtest is not conditional on the existence of at least one corresponding VaR breach and, therefore, this backtest can be applied to all models and across both distributions. The null hypothesis is rejected for all models (both traditional and BN) at the 97.5% confidence level, leading to the conclusion that all models employed do not produce statistically accurate 10-day 97.5% ES forecasts.

Finally, the last backtest we employed was the Du-Escanciano backtest. This backtest's null hypothesis states that the underlying profit and loss distribution observed is the true profit and loss distribution. This backtest is not dependent on the forecasts produced by the various models, but, rather, it tests the return distribution, whether it was assumed or determined by the model. Since the use of BNs in this study produced rolling forecasts that are incorporated into the return PDF, the distribution of returns changed with every forecast, rendering the application of this backtest computationally intensive and of little use (given the daily distribution change). Hence, the Du-Escanciano backtest could only be implemented for the traditional models, where it rejected the null hypothesis for each of the distributions employed in this study at the 97.5% confidence level.

Table 3: Forecasting Error Measures for the 10-day 97.5% Expected Shortfall Forecasts

|  | Normal Distribution | | | Skewed Student's t Distribution | | |
|---|---|---|---|---|---|---|
| **Traditional Models** | **MAE** | **RMSE** | **MAPE** | **MAE** | **RMSE** | **MAPE** |
| *ARCH(1)* | 0.0589 | 0.0766 | 53.779% | 0.0704 | 0.0894 | 62.787% |
| *GARCH(1,1)* | 0.0559 | 0.0717 | 46.842% | 0.0696 | 0.0886 | 61.742% |
| *EGARCH(1,1)* | 0.0546 | 0.0669 | 45.650% | 0.2369 | 0.4394 | 263.921% |
| *RiskMetrics* | 0.0600 | 0.0705 | 51.384% | 0.0687 | 0.0798 | 61.666% |
| **Distribution Agnostic** | **MAE** | **RMSE** | **MAPE** | | | |
| *Historical Simulation* | 0.0917 | 0.0987 | 96.973% | | | |
| *Delta-Normal* | 0.0683 | 0.0720 | 71.551% | | | |
| **BN Models** | **MAE** | **RMSE** | **MAPE** | | | |
| *MMHC* | 0.0918 | 0.0988 | 96.999% | | | |
| *PC (Stable)* | 0.0917 | 0.0987 | 96.974% | | | |
| *SI-HITON-PC* | 0.0917 | 0.0987 | 96.973% | | | |

Note: This table reports the results of the three forecasting error measures, namely the mean absolute error (MAE), the root mean square error (RMSE), and the mean absolute percentage error (MAPE) for the various 10-day expected shortfall (ES) models at the 97.5% confidence level over the period 15 March 1991 to 14 February 2020. The results are based on the logged returns earned on the Standard & Poor's (S&P) 500 index using either the normal or the skewed Student's t distribution as the underlying return distribution (where applicable) and the measures are based on the differences between the forecasted values of the models and the actual returns achieved for each trading day. The total number of forecasts for the study period was 7,286 per model. Note that the historical simulation and the delta-normal models are distribution agnostic, meaning that the forecasted values do not vary with the use of either distribution as the underlying return distribution. The Bayesian network (BN) learning algorithms considered were the max-min hill-climbing (MMHC) algorithm, the Peter and Clark Stable (PC (Stable)) algorithm, and the semi-interleaved HITON parents and children (SI-HITON-PC) algorithm.

The different backtests employed rejected the statistical accuracy of the various models and distributions used to produce the 10-day 97.5% ES forecasts in this study. Hence, we resorted to calculating the models' forecasting efficiencies using the MAE, the RMSE, and the MAPE. Table 3, above, summarizes the results of these forecasting error measures for the various models.

The results show that, overall, the EGARCH model produced the most accurate 10-day 97.5% ES forecasts when using the normal distribution, while the same model, using the skewed Student's t distribution, produced the worst ES forecasts, explaining why it achieved fewer breaches (see Table 2), contradicting the findings of García-Risueño (2025). The historical simulation ranked as the ninth most accurate model, once again highlighting its slow volatility updating abilities (Berkowitz & O'Brien, 2002), being outperformed by most autoregressive models, the exception being the EGARCH model using the skewed Student's t distribution.

The BN models performed similarly to the historical simulation model, despite their superior volatility updating abilities, and ranked as producing the ninth most accurate sets of forecasts, except for the MMHC algorithm, which ranked twelfth most accurate. This, as for the case of the VaR results of Gross, et al., (2025), highlights that the weighting of the one-day-ahead forecast in the calibration period may be insufficient improvement without further refinement.

Examining the effects of the change in distribution for the distribution-dependent models' forecasting accuracies, Table 3 shows that, for all models, the forecasts' accuracies deteriorated with the change from the normal distribution to the skewed Student's t distribution. The most significant deterioration was for the EGARCH model, where the various forecasting error measures' values increased between 334% (MAE) and 557% (RMSE). We ran the results several times to check that no errors were present in the code or any glitches caused these significant changes, and the results were consistent across the various runs. For the other autoregressive models, the deterioration in performance (excluding that of the EGARCH model) ranged from 13% to 32%, showing that, overall, the skewed Student's t distribution produces less accurate tail market risk metrics such as ES, echoing the similar finding of Gross, et al., (2025) for VaR.

The various BNs achieved very similar results. This is expected, as discussed earlier, as each of the PDFs of the BNs only uses a single one-day-ahead forecast when calculating the 10-day 97.5% ES forecasts, while the (much larger) remainder of the calibration period is made up of historical returns. The results are again very similar to those of the historical simulation model and mostly worse (higher scores) than the other traditional models. The differences in the forecasting error measures' values come down to the quality of the forecasts made by the BN algorithms employed. Hence, the BN algorithm that produced the most accurate forecasts is the SI-HITON-PC algorithm, as seen by the lowest forecasting error measure values in Table 3. While the PC (Stable) algorithm scored the same for the MAE and the RMSE (to four decimal places) as the SI-HITON-PC algorithm, the latter scored a marginally lower MAPE score, making it the more accurate algorithm. The MMHC algorithm ranks as the least accurate of the learning algorithms when producing 10-day 97.5% ES forecasts.

### 4.2. Stressed Expected Shortfall

The breaches observed for the various models used to produce the 10-day 97.5% SES forecasts are summarized in Table 4, below. Given the very few breaches observed for the 10-day 97.5% ES forecasts using the same models, it is unsurprising that even fewer breaches were observed for the 10-day 97.5% SES metric, given that this metric is calculated over a stressed period, i.e., the most severe daily profit and loss figures were used to calibrate the models and obtain forecasts. Interestingly, almost all models employed using the skewed Student's t distribution as the underlying return distribution recorded no breaches at all (with the only exception being the ARCH model), while the models experiencing the highest number of breaches were the GARCH model and the EGARCH model (three breaches), both calibrated using the normal distribution as the underlying return distribution. All of the BN algorithms employed produced no breaches at all over the 7,286-trading-day out-of-sample period. This is expected, as the three algorithms produced only three 10-day 97.5% ES breaches each (see Table 1), and, here, the algorithms were used to produce stressed forecasts, resulting in fewer breaches.

Table 4: Number 10-day 97.5% Stressed Expected Shortfall Breaches

| Traditional Models | Normal Distribution | Skewed Student's t Distribution |
|---|---|---|
| *ARCH(1)* | 2 | 1 |
| *GARCH(1,1)* | 3 | 0 |
| *EGARCH(1,1)* | 3 | 0 |
| *RiskMetrics* | 2 | 0 |
| | | |
| *Historical Simulation* | 0 | |
| *Delta-Normal* | 0 | |
| | | |
| **BN Models** | | |
| *MMHC* | 0 | |
| *PC (Stable)* | 0 | |
| *SI-HITON-PC* | 0 | |

Note: This table reports the number of breaches experienced for the various 10-day stressed expected shortfall (SES) forecasting models at the 97.5% confidence level using either the normal distribution or the skewed Student's t distribution as the underlying return distribution (where applicable) using the daily logged returns of the Standard & Poor's (S&P) 500 index from 15 March 1991 to 14 February 2020. A SES breach was recorded where the loss incurred on the S&P 500 index (as measured by its daily logged return) exceeded the forecasted SES figure obtained via one of the models detailed in this table, where the SES figure is calculated over the most severe period preceding the return's date, i.e., over a stressed period. The total number of SES forecasts for the study period was 7,286 per model. Note that the historical simulation and the delta-normal models are distribution agnostic, meaning that the number of breaches experienced does not vary with the use of either distribution as the underlying return distribution. The Bayesian network (BN) learning algorithms considered were the max-min hill-climbing (MMHC) algorithm, the Peter and Clark Stable (PC (Stable)) algorithm, and the semi-interleaved HITON parents and children (SI-HITON-PC) algorithm.

All 10-day 97.5% SES models achieve a 'Green zone' outcome using the regulatory BCBS traffic light test. This is unsurprising, given that the highest number of breaches was three (GARCH and EGARCH, using the normal distribution) across all models and distributions, and the relatively long 7,286-trading-day out-of-sample period. Echoing the finding of Gross, et al., (2025), the BCBS's traffic light model sheds little light on the statistical accuracy of S/ES forecasting models, and it is only useful in deciding whether a high number of breaches observed (if such a number is observed at all) is 'too high', or as a penalty system, as pointed out by Berkowitz, Christoffersen, and Pelletier (2009).

For the conditional backtest, which requires at least one corresponding SVaR breach to assess the statistical accuracy of a model forecasting SES, all but three of the models produced zero SVaR breaches. Those models exhibiting at least one corresponding SVaR breach are the ARCH model, the GARCH model, and the EGARCH model, all of which used the normal distribution as the underlying return distribution. All three models lead to the rejection of the null hypothesis, implying that the forecasts produced are inaccurate. For all other models, including the BNs, the result of 'Cannot perform backtest' is captured in Table 5.

Table 5: Conditional Backtest Results for the 10-day 97.5% Stressed Expected Shortfall Forecasts

| Traditional Models | Normal Distribution | Skewed Student's t Distribution |
|---|---|---|
| *ARCH(1)* | Reject | Cannot perform backtest |
| *GARCH(1,1)* | Reject | Cannot perform backtest |
| *EGARCH(1,1)* | Reject | Cannot perform backtest |
| *RiskMetrics* | Cannot perform backtest | Cannot perform backtest |
| | | |
| *Historical Simulation* | Cannot perform backtest | |
| *Delta-Normal* | Cannot perform backtest | |
| | | |
| **BN Models** | | |
| *MMHC* | Cannot perform backtest | |
| *PC (Stable)* | Cannot perform backtest | |
| *SI-HITON-PC* | Cannot perform backtest | |

Note: This table reports the results of the conditional backtest for banks' internal models based on the number of breaches of various 10-day stressed expected shortfall (SES) models at the 97.5% confidence level using either the normal distribution or the skewed Student's t distribution as the underlying return distribution (where applicable) and the daily logged returns of the Standard & Poor's (S&P) 500 index from 15 March 1991 to 14 February 2020. Note that the historical simulation and the delta-normal models are distribution agnostic, meaning that the number of breaches experienced does not vary with the use of either distribution as the underlying return distribution. The test's null hypothesis states that the SES forecasts observed are the true SES figures. The test can only be carried out if there is at least one stressed value at risk breach. The total number of SES forecasts for the study period was 7,286 per model. The Bayesian network (BN) learning algorithms considered were the max-min hill-climbing (MMHC) algorithm, the Peter and Clark Stable (PC (Stable)) algorithm, and the semi-interleaved HITON parents and children (SI-HITON-PC) algorithm.

Turning to the minimally biased backtest, which does not depend on the existence of at least one corresponding SVaR breach, the null hypothesis is again rejected for all models. Hence, we conclude that none of the models employed (traditional or BN) produced statistically accurate 10-day 97.5% SES forecasts.

As was the case for the ES forecasts, the Du-Escanciano backtest rejected the null hypothesis for each of the distributions employed for the SES forecasts at the 97.5% confidence level.

Given the poor performance of all models across the four backtests employed, we assessed the models' forecasting accuracies using three forecasting errors measures, namely the MAE, the RMSE, and the MAPE, as summarized in Table 6, below.

Table 6: Forecasting Error Measures for the 10-day 97.5% Stressed Expected Shortfall Forecasts

|  | Normal Distribution | | | Skewed Student's t Distribution | | |
|---|---|---|---|---|---|---|
| **Traditional Models** | **MAE** | **RMSE** | **MAPE** | **MAE** | **RMSE** | **MAPE** |
| *ARCH(1)* | 0.0973 | 0.1197 | 107.948% | 0.1115 | 0.1169 | 117.343% |
| *GARCH(1,1)* | 0.0834 | 0.0892 | 88.821% | 0.1337 | 0.1415 | 141.757% |
| *EGARCH(1,1)* | 0.0864 | 0.0935 | 92.717% | 0.1447 | 0.1534 | 150.688% |
| *RiskMetrics* | 0.1363 | 0.1533 | 147.160% | 0.1933 | 0.2105 | 206.258% |
| **Distribution Agnostic** | **MAE** | **RMSE** | **MAPE** | | | |
| *Historical Simulation* | 0.1700 | 0.1728 | 178.391% | | | |
| *Delta-Normal* | 0.1056 | 0.1076 | 110.690% | | | |
| **BN Models** | **MAE** | **RMSE** | **MAPE** | | | |
| *MMHC* | 0.1700 | 0.1728 | 178.409% | | | |
| *PC (Stable)* | 0.1698 | 0.1727 | 178.313% | | | |
| *SI-HITON-PC* | 0.1700 | 0.1728 | 178.391% | | | |

Note: This table reports the results of the three forecasting error measures, namely the mean absolute error (MAE), the root mean square error (RMSE), and the mean absolute percentage error (MAPE) for the various 10-day stressed expected shortfall (SES) models at the 97.5% confidence level over the period 15 March 1991 to 14 February 2020, where the SES figure is calculated over the most severe period preceding the return's date, i.e., over a stressed period. The results are based on the logged returns earned on the Standard & Poor's (S&P) 500 index using either the normal or the skewed Student's t distribution as the underlying return distribution (where applicable) and the measures are based on the differences between the forecasted values of the models and the actual returns achieved for each trading day. The total number of forecasts for the study period was 7,286 per model. Note that the historical simulation and the delta-normal models are distribution agnostic, meaning that the forecasted values do not vary with the use of either distribution as the underlying return distribution. The Bayesian network (BN) learning algorithms considered were the max-min hill-climbing (MMHC) algorithm, the Peter and Clark Stable (PC (Stable)) algorithm, and the semi-interleaved HITON parents and children (SI-HITON-PC) algorithm.

The results in Table 6 show that the most accurate 10-day 97.5% SES forecasts across all models were produced by the GARCH model using the normal distribution as the underlying return distribution, with the EGARCH model using the same distribution being a close second. The RiskMetrics model using the skewed Student's t distribution produced the least-accurate forecasts, while the historical simulation model, the most used internal model by banks, performed only marginally better.

When looking at the changes in forecasting accuracies of the various distribution-dependent models, all models saw a deterioration in forecasting accuracy when using the skewed

Student's t distribution rather than the normal distribution as the underlying return distribution. Again, this highlights the inverse relationship between the number of breaches of a market risk forecasting model and the excess capital held by the banks. The ARCH model's RMSE value marginally decreased when using the skewed Student's t distribution as opposed to the normal distribution, but its MAE and MAPE values increased, indicating an overall deterioration in performance driven by the return distribution change. The GARCH and EGARCH models saw the most significant distribution-change-driven deterioration in forecasting accuracy, averaging around 60% and 65%, respectively. Excluding the marginal improvement in the RMSE value for the ARCH model, all models' forecasts deteriorated in accuracy between 9% and 67% due to the change in distribution. These results highlight the poor tail fit offered by the skewed Student's t distribution relative to the normal distribution when producing 10-day 97.5% SES forecasts. Hence, our results echo the conclusion of Gross, et al., (2025) surrounding the fit of the skewed Student's t distribution and its utility in producing tail metrics for market risk management purposes.

The various BNs employed to produce 10-day 97.5% SES forecasts resulted in relatively similar forecasting error measures across the board. The PC (Stable) learning algorithm scored the lowest across all three measures and, hence, proved to be the most accurate when producing 10-day 97.5% SES forecasts. The remaining algorithms scored marginally higher (i.e., worse) for each of the forecasting error measures, with the second-most accurate algorithm being the SI-HITON-PC algorithm, followed by the MMHC algorithm.

## 5. Conclusion

This study expanded on that of Gross, et al., (2025) on 10-day 99% S/VaR, by applying BNs to 10-day 97.5% ES and SES forecasts. In addition, we provided an extensive assessment of the performances of ten traditional models using either the normal or the skewed Student's t distribution as the underlying return distribution, where the S&P 500 index using data from 15 March 1991 to 14 February 2020 served as a proxy for the equities trading desk of a US bank in the context of the BCBS's Basel accords and banking regulation.

The 7,286 out-of-sample forecasts for each model and each market risk metric were backtested using the BCBS's traffic light test, the conditional and minimally biased backtests of Acerbi and Székely (2014, 2017), and the Du-Escanciano (2017) backtest was used to statistically evaluate the fits of the normal and skewed Student's t distributions to the true profit and loss distribution for the traditional models. The BCBS's traffic light test resulted in a 'Green zone' outcome for all models. Echoing the findings of Gross, et al., (2025), we conclude that this regulatory test is not a statistical backtest of forecasting accuracy but, rather, a test for excessive breaches, offering little practical use beyond a regulatory requirement. The results for the conditional backtest, where it could be carried out (as it depends on the existence of at least one corresponding S/VaR breach), as well as the minimally biased backtest (for which such dependence does not exist), were the rejections of the null hypotheses for all applicable models, indicating that the relevant S/ES forecasts were statistically inaccurate at the 97.5% confidence level. Finally, the Du-Escanciano backtest rejected the fit of both the normal and skewed Student's t distributions as the true underlying profit and loss distribution of the equities trading desk of a US bank, proxied by the S&P 500 index return distribution, also at the 97.5% confidence level.

Due to the limited use of backtesting results (either a uniform 'Green zone' result or rejected null hypotheses), we turned to assess model accuracy by examining the forecasting accuracies of models, through the comparison of their forecasts with the true returns earned on the S&P 500 index. For the 10-day 97.5% ES forecasts, the most accurate model was the EGARCH(1,1) using the normal distribution as the underlying return distribution, while the same was true when using the GARCH(1,1) model and the normal distribution. Our results also highlight the poor fit of the skewed Student's t distribution when it comes to assessing and quantifying tail risk, as evidenced by the deterioration of the forecasting accuracies of the autoregressive models when using the skewed Student's t distribution instead of the normal distribution as the underlying return distribution (between 9% and 67%). This poorer fit was also found to be true for the 10-day 97.5% ES forecasts, where the forecasting accuracies of the autoregressive models deteriorated by between 13% and 557% (or 13% and 32% once we exclude the deterioration in EGARCH model's forecasting accuracy).

Turning our attention to the three BNs, they, too, yielded very few breaches when producing 10-day 97.5% ES forecasts and no breaches when producing 10-day 97.5% SES forecasts. As for the traditional models, backtesting the BNs using the BCBS's traffic light test resulted in a 'Green zone' outcome for all three algorithms, while the conditional and minimally biased backtests of Acerbi and Székely (2014, 2017) could not be performed and resulted in the rejection of the null hypothesis, respectively. The latter means that the BNs, just like the traditional models, produced statistically inaccurate S/ES forecasts at the 97.5% confidence level. The forecasting error measures used indicated that the SI-HITON-PC algorithm produced the most accurate 10-day 97.5% ES forecasts, while, for the 10-day 97.5% SES forecasts, it was the PC (Stable) algorithm that proved most accurate, where both sets of results were comparable to those achieved when using traditional models, as was the case for S/VaR in Gross, et al., (2025).

This study expanded the literature in several meaningful ways. First, we provided a comprehensive assessment of the performances of several traditional models when producing 10-day 97.5% S/ES forecasts in the context of banking regulation and the BCBS's Basel accords, an area which saw little research since the BCBS's switch to use ES as its primary market risk metric in the Basel III accord. This comparison was performed over an extended out-of-sample period (1991 to 2020) using two distribution assumptions (where applicable), four backtests, and three forecasting error measures, providing a robust and comprehensive assessment of model performance. We found that while both distributions' fits are rejected by the Du-Escanciano (2017) backtest as the true underlying return distribution, the normal distribution's tail fit produced more accurate tail market risk forecasts than the skewed Student's t distribution, despite the assumed superior distributional overall fit of the latter. Finally, we extended the application of DBNs beyond S/VaR to S/ES, or the rest of the tail of the return distribution, extending the analyses and conclusions presented by Gross, et al., (2025).

The minimal improvement in forecasting accuracies of the BNs relative to simpler models such as the historical simulation is driven primarily by the networks' one-part contribution to the return PDF, being the one-day-ahead forecasted return of the S&P 500 index, relative to the remaining 1,263 parts, being the historical returns observed in the past. Hence, even though a BN incorporates a forward-looking methodology, which should, in theory, yield more accurate market risk metrics, the equal weighting of this BN-derived one-day-ahead return forecast relative to the rest of the return distribution has minimal impact in practice. Hence, future research should focus on assigning more weight to the BN's contribution to the return PDF and assessing such methodology's potential improvement in forecasting accuracy for VaR, ES, and their stressed counterparts.

# Appendix A: Variables used to train the Bayesian networks

Table 7: Variables used to train the Bayesian networks

| Variable Name | Classification |
|---|---|
| *S&P 500 index (closing value)* | Target variable |
| *Australian Dollar/US Dollar* | Currency Exchange Rate |
| *Bloomberg Commodities Index* | Financial |
| *Bloomberg US Treasury Total Return Index* | Financial |
| *Canadian Dollar/US Dollar* | Currency Exchange Rate |
| *Canola Price* | Commodity |
| *Corn Price* | Commodity |
| *Euro/US Dollar* | Currency Exchange Rate |
| *Federal Reserve Lending Rate* | Economic |
| *Financial Times Stock Exchange 100 Index* | Financial |
| *Gold Price* | Commodity |
| *Great British Pound/US Dollar* | Currency Exchange Rate |
| *Japanese Yen/US Dollar* | Currency Exchange Rate |
| *Nikkei Index* | Financial |
| *Oats Price* | Commodity |
| *Russell 1000 Index* | Financial |
| *Russell 2000 Index* | Financial |
| *Russell 3000 Index* | Financial |
| *Silver Price* | Commodity |
| *Soybean Meal Price* | Commodity |
| *Soybean Oil Price* | Commodity |
| *Soybean Price* | Commodity |
| *S&P Australian Stock Exchange Index* | Financial |
| *S&P Toronto Stock Exchange Index* | Financial |
| *Three-month US London Interbank Offer Rate* | Financial |
| *Topix Index* | Financial |
| *US Consumer Price Index* | Economic |
| *US Corporate Aaa-rated Ten-year Spread* | Financial |
| *US Corporate Bonds Index* | Financial |
| *US Disposable Income Growth Index* | Economic |
| *US Full Employment* | Economic |
| *US Jobless Claims* | Economic |
| *US M1 Money Supply* | Economic |
| *US M2 Money Supply* | Economic |
| *US s Manufacturing Index* | Economic |
| *US Manufacturing Tendency Index* | Economic |
| *US Non-Farm Payroll* | Economic |
| *US Part Time Employment* | Economic |
| *US Policy Uncertainty Index* | Political |
| *West Texas Intermediate Price* | Commodity |
| *Wheat Price* | Commodity |

Note: This table lists the variables used to train the Bayesian networks (BNs) using the various learning algorithms in this study. The data for the variables were obtained from the Bloomberg database over the period 15 March 1991 to 14 February 2020, as well as for a calibration period and a training period preceding this start date. The data relate to variables from the United States (US) as identified (from the literature or otherwise) to causally relate to the target variable, the Standard & Poor's (S&P) 500 index.